# The Leiden/Argentine/Bonn (LAB) Survey of Galactic HI:

## Final data release of the combined LDS and IAR surveys with improved stray-radiation corrections

P. M. W. Kalberla[1], W. B. Burton[2,3], Dap Hartmann[2,4], E. M. Arnal[5,6]*, E. Bajaja[5]*, R. Morras[5,6]*, and W. G. L. Pöppel[5]*

[1] Radioastronomisches Institut der Universität Bonn, Auf dem Hügel 71, 53121 Bonn, Germany
   e-mail: pkalberla@astro.uni-bonn.de
[2] Sterrewacht Leiden, Postbus 9513, Leiden, 2300RA, Netherlands
[3] National Radio Astronomy Observatory, 520 Edgemont Road, Charlottesville, Virginia 22903, U.S.A.
   e-mail: bburton@nrao.edu
[4] Delft University of Technology, Postbus 5, Delft, 2600AA, Netherlands
   e-mail: L.Hartmann@TBM.TUDelft.nl
[5] Instituto Argentino de Radioastronomía, c.c. 5, (1894) Villa Elisa
   e-mail: bajaja@iar.unlp.edu.ar, arnal@iar.unlp.edu.ar, rimorras@isis.unlp.edu.ar,
   wpoeppel@iar.unlp.edu.ar
[6] Facultad de Ciencias Astronómicas y Geofísicas, UNLP, Av. Paseo del Bosque S/N, 1900 La Plata, Argentina

**Abstract.** We present the final data release of observations of $\lambda$21-cm emission from Galactic neutral hydrogen over the entire sky, merging the Leiden/Dwingeloo Survey (LDS: Hartmann & Burton, 1997) of the sky north of $\delta = -30°$ with the Instituto Argentino de Radioastronomía Survey (IAR: Arnal et al., 2000, and Bajaja et al., 2005) of the sky south of $\delta = -25°$. The angular resolution of the combined material is HPBW $\sim 0°.6$. The LSR velocity coverage spans the interval $-450$ km s$^{-1}$ to $+400$ km s$^{-1}$, at a resolution of 1.3 km s$^{-1}$. The data were corrected for stray radiation at the Institute for Radioastronomy of the University of Bonn, refining the original correction applied to the LDS. The rms brightness-temperature noise of the merged database is $0.07 - 0.09$ K. Residual errors in the profile wings due to defects in the correction for stray radiation are for most of the data below a level of $20 - 40$ mK. It would be necessary to construct a telescope with a main beam efficiency of $\eta_{MB} \gtrsim 99\%$ to achieve the same accuracy. The merged and refined material entering the LAB Survey of Galactic HI is intended to be a general resource useful to a wide range of studies of the physical and structural characteristics of the Galactic interstellar environment. The LAB Survey is the most sensitive Milky Way HI survey to date, with the most extensive coverage both spatially and kinematically.

**Key words.** surveys – ISM: general – Galaxy: general radio lines: ISM – ISM: clouds – ISM: atoms – ISM: HI

## 1. Introduction

Hydrogen is the principal observed component of the interstellar medium in our Galaxy. Observations of emission from hydrogen in its neutral atomic state continue to provide insights into differing temperature and density regimes of the interstellar medium, and into the kinematic and morphological properties of the Milky Way. The $\lambda$21-cm line is a particularly useful tracer because of the ubiquity of HI – no direction on the sky has ever been found devoid of quite easily observable emission – and because under most circumstances the interstellar medium is transparent at the 21-cm wavelength, revealing the entire Galaxy, well beyond the optical horizon.

The important observational parameters relevant for large-scale surveys of Galactic HI include the extent and resolution of the sky coverage, the extent and resolution of the coverage in velocity, and the sensitivity to weak emission. For the extended northern Milky Way sky at $\delta > -30°$, the Leiden/Dwingeloo Survey published by Hartmann & Burton (1997) improved upon the earlier Hat Creek 85-foot survey at $|b| < 10°$ of Weaver & Williams (1973) in terms of sensitivity and sky coverage, upon the Hat Creek survey at higher northern latitudes of Heiles & Habing (1974) in terms of sensitivity and velocity coverage, and upon the Bell Labs Survey of Stark et al. (1992) in terms of velocity resolution and spatial resolution. Although the LDS had been corrected for most contamination by stray radiation by Hartmann, Kalberla, Burton, & Mebold (1996), no correction for reflected ground radiation was applied because of various uncertainties and because of limitations of computer power. These uncertainties have since been resolved, computer power is now sufficient, and the algorithms for stray-radiation correction have been refined: the LDS material as used here represents an improved second edition of the data.

**Table 1.** Observational parameters of the two H I surveys combined, and further corrected, to form the LAB Survey of Galactic H I

| Parameter | IAR / Villa Elisa | Leiden/Dwingeloo |
|---|---|---|
| Antenna diameter | 30 m | 25 m |
| Main beam FWHM | $30\rlap{.}'0$ | $35\rlap{.}'7$ |
| Mounting | equatorial | alt-az |
| Declination range | $< -25°$ | $> -30°$ |
| System temperature | $\sim 35$ K | $\sim 35$ K |
| Front end | HEMT He-cooled | FET He-cooled |
| RMS noise | $\sim 0.07$ K | $\sim 0.09$ K |
| Back end | 1008 channel DAC | 1024 channel DAC |
| Bandwidth | 5 MHz | 5 MHz |
| Velocity coverage | $-450$ to $+450$ km s$^{-1}$ | $-450$ to $+400$ km s$^{-1}$ |
| Velocity resolution | 1.27 km s$^{-1}$ | 1.25 km s$^{-1}$ |
| Channel separation | 1.05 km s$^{-1}$ | 1.03 km s$^{-1}$ |
| IAU standard regions | S9 | S8 and S7 |
| Observing dates | 1994 to 2000 | 1989 to 1993 |
| Number of spectra | 50980 | 138830 |
| Grid in $l, b$ | $0\rlap{.}°5/\cos(b), 0\rlap{.}°5$ | $0\rlap{.}°5/\cos(b), 0\rlap{.}°5$ |

For the southern Milky Way sky at $\delta < -25°$, the IAR survey published by Arnal et al. (2000) and Bajaja et al. (2005) similarly improves upon the Parkes 60-foot survey at $|b| < 10°$ of Kerr et al. (1986) and, at more southern latitudes, of Cleary, Haslam, & Heiles (1979), and upon the IAR survey of Colomb, Pöppel, & Heiles (1980).

The principal challenge facing the LDS and IAR efforts lay in the correction for stray radiation. The earlier surveys were not similarly challenged, because the sensitivies reached did not approach the levels where contamination from stray radiation was particularly important. The Bell Labs Survey (Stark et al., 1992), having been made with a horn antenna, suffers significantly less from stray radiation than the other efforts, all made with blocked-aperture antennae whose feed-support structures reflected unwanted emission from the general H I sky. The Bell Labs Survey is, however, characterized by a coarse spatial resolution (2°) as well as by some residual instrumental problems. We describe here a refined correction for stray radiation which results in an improved second edition of the Leiden/Dwingeloo Survey; the final data release of the Instituto Argentino de Radioastronomía IAR Survey, published by Bajaja et al. (2005), incorporates a similar stray-radiation correction. Both surveys have now been combined to a provide the LAB Survey, a coherent database representing the entire sky at unprecedented sensitivity and coverage in space as well as velocity.

## 2. Parameters of the all-sky LAB Survey merging the LDS and IAR data

The instrumental parameters, the observational protocols, and the methods of data reduction for the LDS and the IAR are described by Hartmann & Burton (1997) and Arnal et al. (2000), respectively. The principal observational parameters for the surveys are summarized in Table 1. The entries in this table show that the specifications for both surveys closely match each other. In important regards, the parameters represent an order of magnitude improvement with respect to previous surveys. The brightness-temperature sensitivity is $0.07 \lesssim \sigma_{\mathrm{rms}} \lesssim 0.09$ K. The velocity resolution is $\sim 1$ km s$^{-1}$, over a velocity range extending from $-450$ km s$^{-1}$ to $+400$ km s$^{-1}$. We demonstrate in Sect. 4 that the contributions from stray radiation have been decreased by an order of magnitude.

A principal limitation of the LDS and IAR efforts lies in the moderate spatial resolution and an observing grid that violates Nyquist sampling. This limitation was in both cases unavoidable due to the moderate size of the antennae, to the relatively high-noise receivers, to the single-feed arrangement, and to practical constraints on the length of time which could be devoted to the projects. The FWHM beam is about $0\rlap{.}°5$; the observational grid is similar for both surveys, namely $0\rlap{.}°5$ in latitude and $\sim 0\rlap{.}°5/\cos(b)$ in longitude.

## 3. Data reduction and calibration

Details about the data reduction and calibration have been given by Hartmann & Burton (1997) and Hartmann, Kalberla, Burton, & Mebold (1996), for the Leiden/Dwingeloo Survey, and by Arnal et al. (2000), and Bajaja et al. (2005), for the IAR Survey. All of the data reduction and calibration procedures were carried out entirely independently for each of the surveys, although in general similar, quite standard, routines were followed. The first step involved the gain and bandpass calibration. Afterwards the contribution of the stray radiation was calculated and subtracted, and finally the instrumental baseline was removed. Both sets of observations suffered significantly from radio-frequency interference. Since it was not possible to identify all of the cases of interference in time to re-observe the affected positions, special care was necessary to recognize and then remove such spurious signals after the calibration.

A major difference between the Dwingeloo and the Villa Elisa telescopes lies in the mountings. The Dwingeloo telescope has an alt-azimuth mounting with an elevation limit of 5°. The IAU standard positions S7 and S8 (Williams, 1973) were used to monitor gain fluctuations and to gauge the conversion of antenna temperature to brightness temperature. The Villa Elisa telescope has an equatorial mounting with a lowest elevation of $34°\!.8$. Due to limitations in hour angle tracking it was not possible to observe the primary calibrator source S9 all the time. Therefore 10 additional calibrators were used for calibration of the IAR survey, as described by (Bajaja et al., 2005). Despite the differences in the calibrations, the two surveys agree well in the 5° declination range of overlap.

### 3.1. Correction for stray radiation

Both surveys have been corrected for stray radiation by convolving the observable sky with the antenna diagram of the respective telescope. Such a calculation demands a priori knowledge of the stray pattern, $(SP)$, of the antenna diagram, $P$, and of the true temperature distribution, $T$, on the sky. For a telescope with a main beam efficiency $\eta_{MB}$, the brightness temperature $T_B$ can be derived from the observed antenna temperature $T_a$. This procedure is described in more detail by Kalberla, Mebold, & Reich (1980); Kalberla, Mebold, & Velden (1980), Hartmann, Kalberla, Burton, & Mebold (1996), and Bajaja et al. (2005). The brightness temperature is then given by

$$T_B(x, y) = \frac{T_a(x, y)}{\eta_{MB}} - \frac{1}{\eta_{MB}} \int_{SP} P(x - x', y - y') T(x', y') dx' dy'. \tag{1}$$

This is in principle a simple calculation; however, a priori neither $T_B$ (substituting $T$) nor $P$ is known sufficiently well. The way out involves the so-called resolving kernel method (Kalberla, 1978), possible for all telescopes with a main beam efficiency $\eta_{MB} > 0.5$. This deconvolution is based on the observed antenna temperatures, $T_a$, while $P$ is replaced by the resolving kernel $Q$ which can be derived iteratively from $P$. Then

$$T_B(x, y) = \frac{T_a(x, y)}{\eta_{MB}} - \frac{1}{\eta_{MB}} \int_{SP} Q(x - x', y - y') T_a(x', y') dx' dy'. \tag{2}$$

Since nearly all radio telescopes have main beam efficiencies of $\eta_{MB} \gtrsim 0.7$, the observations are thus, generally, correctable for stray radiation. The resolving kernel method was used by Kalberla, Mebold, & Reich (1980); Kalberla, Mebold, & Velden (1980) for the Effelsberg telescope, and later by Hartmann, Kalberla, Burton, & Mebold (1996) for the Dwingeloo telescope. The first calculations in 1980 were based on a set of 86 profiles at 10 positions; the calculation for each of the profiles demanded 15 minutes of CPU time. Since it turned out that the antenna pattern $P$ is not known well, it has been necessary to model some of the major features like stray cones and spillover. Eq. 2 was solved by varying beam parameters in such a way that a consistent solution could be obtained. It is obvious that the accuracy of such a procedure is CPU limited. For the correction of the LDS, as done by Hartmann et al. some 17 years later, a few hundred profiles were used for parameter fitting; the correction of the total LDS survey took a CPU week on a DEC-Alpha processor. This version of the correction is the one represented by the original edition of the LDS published in the Hartmann & Burton (1997) "Atlas".

Advances in computer technology are now such that 1000 profiles per minute can be corrected using a standard PC. This allows us to *use the total observed database* for the necessary fitting of the antenna parameters. Furthermore, the correction can easily be reiterated. A direct solution according to Eq. 1 is possible as soon as the total survey has been converted to brightness temperatures for the first time (see also Bajaja et al., 2005 for discussion). The solution according to Eq. 1 could be improved if our knowledge about the beam pattern were better. Measurements like those made by Hartsuijker et al. (1972) would be needed, but this is no longer an option for either the Dwingeloo or the Villa Elisa telescope.

### 3.2. Second edition of the LDS, incorporating a refined stray-radiation correction

A detailed description of stray radiation effects contaminating the LDS data was given by Hartmann, Kalberla, Burton, & Mebold (1996). At the time this survey was published by Hartmann & Burton (1997), it was already noted that the data still suffered from residual low-level stray radiation, principally due to reflections from the ground (see Fig. 24 of Hartmann, Kalberla, Burton, & Mebold, 1996).

The entire LDS database was corrected a second time, making use of several improvements. Thus, for example, the input sky could now include the data from the IAR southern survey; in addition, the velocity range of the input sky was extended to include also stray radiation from the Magellanic System ($-263 < v < +385$ km s$^{-1}$). The correction was calculated according to Eq. 1, using brightness temperatures for the deconvolution. Concerning the antenna diagram, the kernel $Q$ was replaced by $P$; a correction for ground reflections (Kalberla et al., 1998) was included; and the parameters for the spillover range were determined once more with better accuracy. Some additional improvements were applied to the code.

As the final step in the data reduction, we corrected the instrumental baseline. The procedure used was identical to the one described by Bajaja et al. (2005) for the IAR survey, except that some additional channels lying near the edge of the bandpass have been excluded from the process. We determined the baselines by fitting, successively, polynomials of order 1 to 4, each time searching for possible profile components which need to be disregarded in fitting. After applying the 4$^{th}$ order polynomial we removed standing waves by applying a sine wave fit, allowing a fit of two sine waves simultaneously. As the last step the 4$^{th}$ order polynomial fit was repeated, using exactly the same parameters as before.

Many of the positions in the LDS had been observed more than once, in the process of accumulating integrations of five minutes per direction. When preparing the final database, the criteria to select the best profile out of a number of multiple observations from the LDS were revised. We used a Gaussian decomposition for this purpose (Haud, 2000): among multiple profiles those were selected that according to their Gaussian components agreed appropriately with their neighborhood.

### 3.3. Internal consistency

A good way to detect problems with observational data is to compare observations from different telescopes. Higgs & Tapping (2000) compared Dwingeloo data with data which they had observed using the 26-m DRAO telescope and pointed out, correctly, that some of the LDS positions suffer from evident calibration problems. They present in their Table 3 a list of 48 positions in the LDS with suspected error; the LDS spectra in the tabulated directions differ by some 10%, or in a few cases even more, in either integrated intensity or in peak intensity, from the observations Higgs & Topping made with the DRAO telescope. Unfortunately, none of these positions was among those in the LDS with multiple spectra, and so we were unable to correct them by eliminating faulty partial integrations. These faulty spectra remain in the refined second edition. To a high probability most of these problems will have been caused by interference, either affecting the temperature scale or the profile shape with spurious components. There may be other such distorted spectra amongst the 138830 profiles represented in the second edition of the LDS data, in addition to those pointed out by Higgs & Tapping; they can probably best be recognized by unacceptably sharp deviations from neighboring positions, because of the expected short duration of the interference. Profiles at a few positions were also found to deviate significantly in shape; Higgs & Tapping suggest that these deviations are due to problems in correcting for stray radiation. In case of an incorrect calibration the profile shape may be affected by the correction algorithm since the subtraction of stray radiation demand a correct calibration. But we failed to verify such problems in the calculated stray radiation; indeed, in general, for small changes in position, the stray radiation profile varies little, provided that the observations are performed at the same time. Strong variability, however, is found if the date or season of the observations changes a lot. In particular, spurious emission originating far away from the main beam may cause artifacts in channel maps (see Fig. 1 & 2). These features are patchy in the case of LDS since $5° \times 5°$ fields have been mapped, and stripy when scanning at constant declination, as pertains in the case of the Bell Labs survey.

The LDS and the IAR surveys overlap within the declination range $-30° < \delta < -25°$. Data in the region of overlap have been used for consistency checking by Bajaja et al. (2005, Fig. 3). Both surveys, reduced and calibrated independently, agree quite well. Still, a note of caution is appropriate. Positions that were observed only once can obviously not be checked for internal consistency. The Dwingeloo telescope no longer functions, so nothing can be done, with that instrument in any case, to repair defective profiles.

Bajaja et al. (2005) point out that also in case of the IAR survey there are some residual problems due to interference, which will predominantly affect the baselines at low levels. Such problems would be relatively easy to identify since they would be expected to be correlated over a $5 \times 5$ position observing grid, and to show up as patchy features in the channel maps. Approximately 1 or 2% of the positions may be affected in this way, mostly at a very low intensity.

## 4. Availability of the LAB Survey in digital form

We combined the refined edition of the LDS and the IAR survey within a single FITS data cube. In the first step, we used a spline interpolation (Press, Teukolsky, Vetterling, & Flannery, 1992) to convert the 1008-channel IAR data to the 1024-channel velocity grid of the LDS (see Table 1). For a spatial regridding, a Gaussian interpolation function with a FWHM of 18.′0 was used. This results in a slight degradation in spatial resolution: from 30.′0 to 35.′0 in the case of the IAR survey, and from 35.′7 to 40.′0 in the case of the LDS. Profiles within the overlap region at $-27° < \delta < -25°$ [1] were averaged with equal weight, leading approximately to a resolution of $37\rlap{.}'5$. Please note that the data are not Nyquist sampled, the effective resolution depends therefore also on the observed grid.

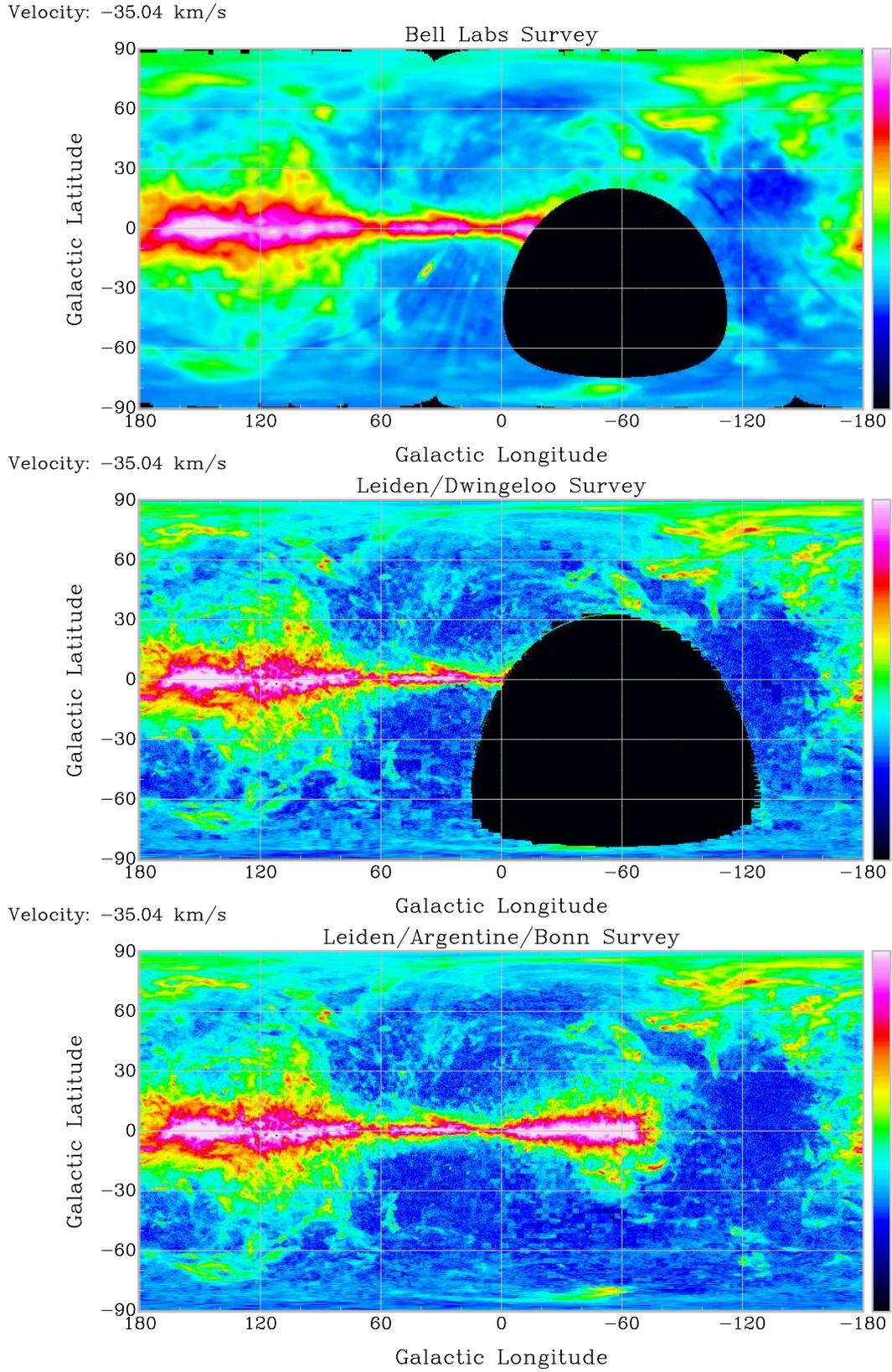

**Fig. 1.** Images of the H I sky contributed by emission in single channels centered near $v = -35 \, \mathrm{km \, s^{-1}}$. *top:* H I emission in a single channel, $5 \, \mathrm{km \, s^{-1}}$ wide, from data in the Bell Labs Survey (Stark et al., 1992) . *middle:* H I emission in a single channel, $1.3 \, \mathrm{km \, s^{-1}}$ wide, from data in the first edition of the LDS (Hartmann & Burton, 1997). *bottom:* H I emission in a single channel, $1.3 \, \mathrm{km \, s^{-1}}$ wide, from the LAB Survey of the entire Galactic sky, combining the IAR and LDS material and incorporating a refined determination of the stray-radiation properties. The intensity scale is logarithmic, emphasizing low level emission ($-0.21 < T < 50 \, \mathrm{K}$), and is the same in all maps of Figs. 1&2.

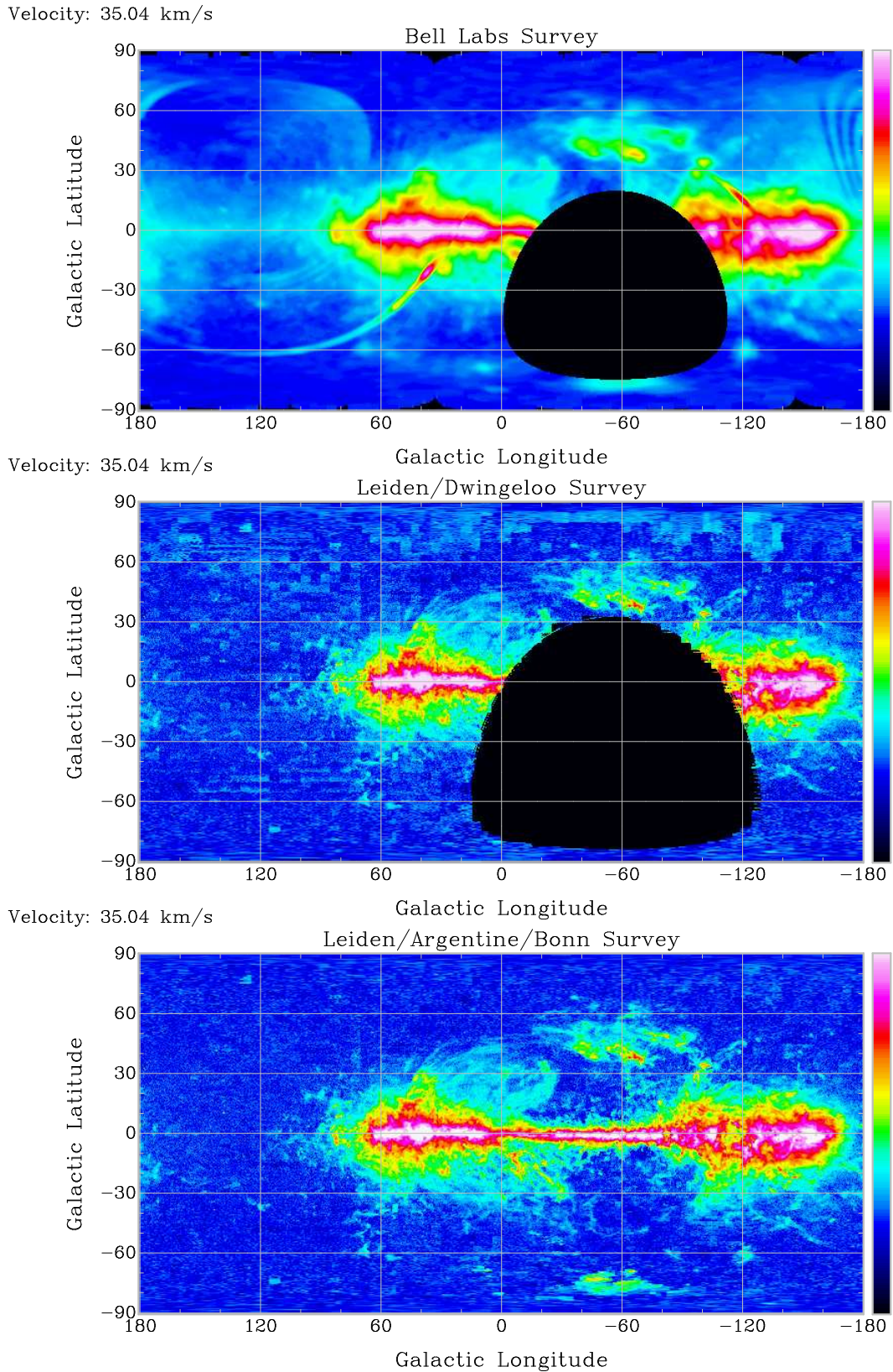

**Fig. 2.** Images of the H ɪ sky contributed by emission in single channels centered near $v = +35\,\mathrm{km\,s^{-1}}$. *top:* H ɪ emission in a single channel, $5\,\mathrm{km\,s^{-1}}$ wide, from data in the Bell Labs Survey (Stark et al., 1992) . *middle:* H ɪ emission in a single channel, $1.3\,\mathrm{km\,s^{-1}}$ wide, from data in the first edition of the LDS (Hartmann & Burton, 1997). *bottom:* H ɪ emission in a single channel, $1.3\,\mathrm{km\,s^{-1}}$ wide, from the LAB Survey of the entire Galactic sky, combining the IAR and LDS material and incorporating a refined determination of the stray-radiation properties. The intensity scale is logarithmic, emphasizing low level emission ($-0.21 < T < 50$ K), and is the same in all maps of Figs. 1&2.

A visual inspection of such a FITS cube shows some annoying flicker when running video loops. A DC offset in the Dwingeloo correlator causes a correlated channel-to-channel noise. An easy cure for such a defect is to apply a Hanning smoothing to the whole database. We provide such a smoothed database, with an effective velocity resolution of the FITS maps in this case being $1.9\ \mathrm{km\,s^{-1}}$. For quality control of the database concerning stray radiation we provide, in addition, a survey version which includes this antenna response. The FITS cubes containing the LAB Survey data are available at the CDS. (********** give link to catalog VIII/76 here ************).

The CDS site also gives a table of H I column densities, calculated under the common assumption of negligible optical depth and recorded on the half-degree lattice points of the LAB Survey for several different characteristic velocity intervals in the range $-25 < v < +25\ \mathrm{km\,s^{-1}}$ to $-400 < v < +400\ \mathrm{km\,s^{-1}}$.

## 5. Sample maps displaying the LAB Survey

We give in Figures 1 and 2 sample channel maps which illustrate the refinements to the LDS in its second edition, and which illustrate the sensitivity of the LAB Survey compared to that of the Bell Labs Survey.

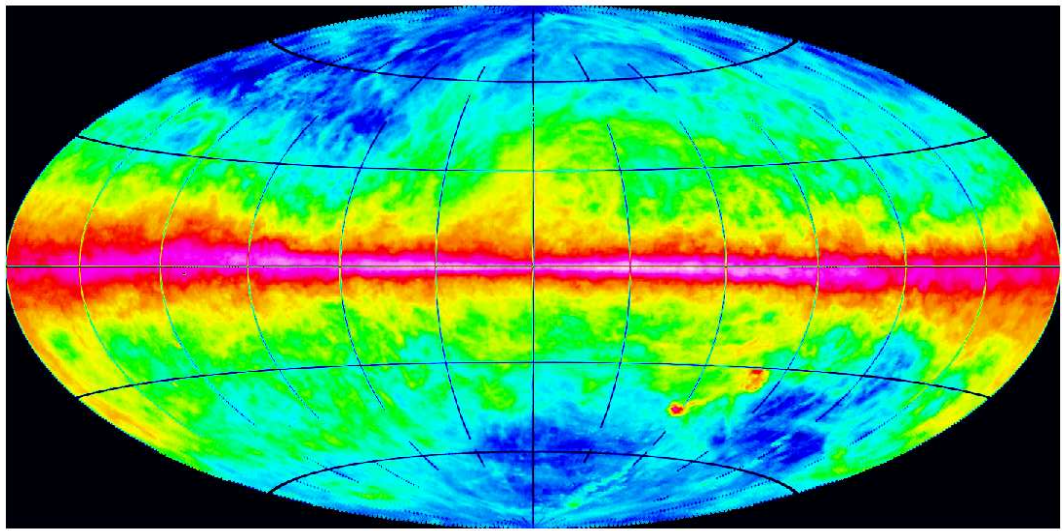

**Fig. 3.** H I emission integrated over the velocity range $-400 < v < +400\ \mathrm{km\,s^{-1}}$ in the LAB dataset, shown in an Aitoff projection. The Galactic center is in the middle. The integrated emission ($0 < N_H < 2\ 10^{22}\ \mathrm{cm^{-2}}$, logarithmic scale) yields column densities under the assumption of optical transparency; this assumption may be violated at latitudes within about $10°$ of the Galactic equator.

We give in Figures 3, 4, and 5 sample maps, in Aitoff projection, showing the integrated H I emission over three representative velocity ranges. Figure 3 shows the total H I column density, integrated over the velocity range $-400 < v < +400\ \mathrm{km\,s^{-1}}$. Figures 4 and 5 show the integrated emission over two velocity ranges ($-400 < v < -100\ \mathrm{km\,s^{-1}}$ and $100 < v < +400\ \mathrm{km\,s^{-1}}$, respectively) which represent H I gas mostly removed, kinematically as well as spatially, from the conventional Galactic gaseous disk. The maps illustrate, schematically, that the combined surveys merge smoothly, and that the all-sky coverage afforded by the LAB Survey is important for study of many extended features.

Figure 1 shows the channel maps at $v = -35\ \mathrm{km\,s^{-1}}$; Figure 2 shows the channel maps at $v = +35\ \mathrm{km\,s^{-1}}$. The top panel of both of these figures shows a channel map from the Bell Labs Survey, derived from the spectra published by Stark et al. (1992, Appendix A). For the regridding we applied a Gaussian smoothing function of $2°$ since we wanted to avoid the heavy smoothing as applied by Stark et al. (1992, Appendix B) to their published maps. No attempt was made to clean these data for sidelobe effects, such as visible at $l = 36\rlap{.}°5$, $b = -21°$, for example, and elsewhere. The middle panel of both figures shows a channel map based on the LDS data as published by Hartmann & Burton (1997). The bottom panel shows a channel map from the LAB Survey, made using the Hanning-smoothed data cube.

---

[1] LDS spectra for $\delta \lesssim -27°$ were in general observed at very low elevations. Such profiles are significantly affected by ground radiation. To avoid a degradation of the fits database by baseline defects we excluded such data.

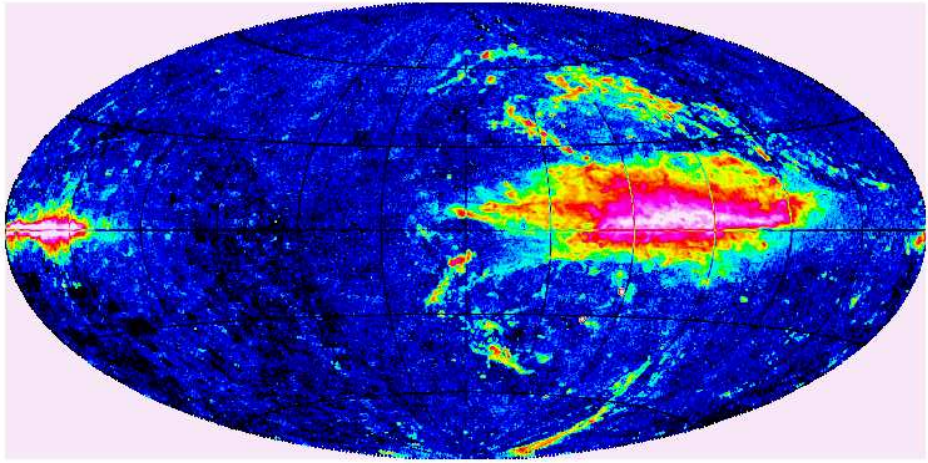

**Fig. 4.** Anticenter view of the Hɪ emission integrated over the velocity range $-400 < v < -100$ km s$^{-1}$ in the LAB dataset, shown in an Aitoff projection. The integrated emission ($0 < N_H < 5\ 10^{20}$ cm$^{-2}$, logarithmic scale) yields column densities under the assumption of optical transparency; this assumption may be violated at latitudes within about $10°$ of the Galactic equator.

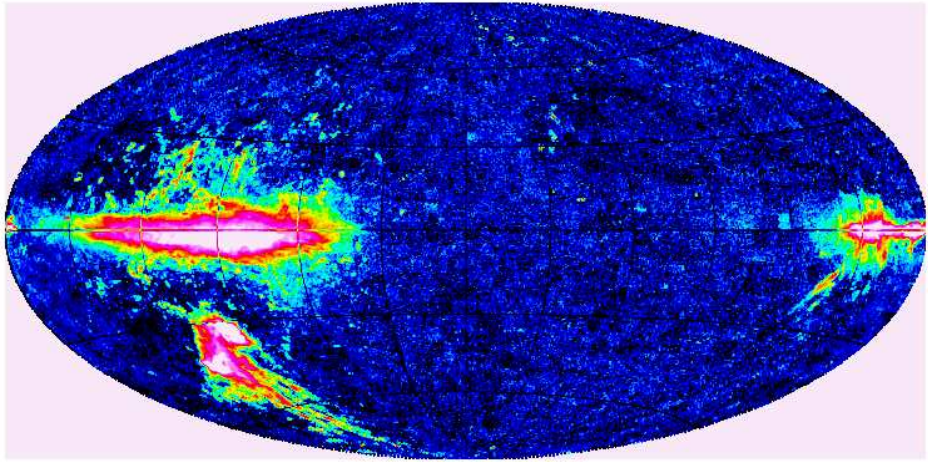

**Fig. 5.** Anticenter view of the Hɪ emission integrated over the velocity range $+100 < v < +400$ km s$^{-1}$ in the LAB dataset, shown in an Aitoff projection. The integrated emission ($0 < N_H < 5\ 10^{20}$ cm$^{-2}$, logarithmic scale) yields column densities under the assumption of optical transparency; this assumption may be violated at latitudes within about $10°$ of the Galactic equator.

Figures 1 and 2 demonstrate that the data entering the LAB Survey are improved with respect to those of the first edition of the LDS. It is also clear that the LAB Survey improves upon the Bell Labs material not only in terms of sky coverage and angular resolution, but also in terms of sensitivity and freedom from residual spurious effects.

The Bell Labs data shown in the upper panels of Figures 1 and 2 indicate some obvious effects caused by stray radiation. Most of these are at a level of 200 to 400 mK, e.g. at the level of about 300 mK near $l = 180°$, $b = 60°$, at about 400 mK near $l = 150°$, $b = 0°$, and at about 200 mK near $l = 200°$, $b = 40°$. We believe that for the major part of the LAB combined material similar problems are below a level of 20 to 40 mK, an order of magnitude less. At such a level it is difficult to distinguish stray radiation from baseline defects. This comparison may be used to estimate the efficiency of the correction algorithm. The Bell Labs horn-reflector antenna has a sidelobe efficiency of 8%. Correspondingly, the contribution in the LAB combined survey is that of an equivalent telescope with a sidelobe efficiency of 0.8%. A similar guess is possible if one considers Hɪ column densities. As shown by Bajaja et al. (2005, Figs. 3 & 4), systematic errors of the LAB database are probably at a level a factor ten below that of the systematic errors in the Bell Labs Survey. We conclude that the material constituting the Leiden/Argentine/Bonn Survey is equivalent to a dataset which would be observable with a telescope having a main beam efficiency of $\eta_{MB} \gtrsim 0.99$.

*Acknowledgements.* This work was partially financed by the German Research Foundation (Deutsche Forschungsgemeinschaft) DFG project number Ka 1265/2–1, also by the Consejo Nacional de Investigaciones Cinetíficas y Técnicas (CONICET) of Argentina under Project PIP 2277/00. We are grateful to Ulrich Mebold for continuous support and to Urmas Haud for running his Gaussian analysis on the survey database for the sake of quality control.